# Strain-induced exciton decomposition and anisotropic lifetime modulation in a GaAs micromechanical resonator

Ryuichi Ohta[1], Hajime Okamoto[1], Takehiko Tawara[1,2], Hideki Gotoh[1], Hiroshi Yamaguchi[1]

[1] *NTT Basic Research Laboratories, NTT Corporation,*
[2] *NTT Nanophotonics Center, NTT Corporation*
*3-1 Morinosato Wakamiya, Atsugi-shi, Kanagawa 243-0198, Japan*

We demonstrate mechanical modulations of the exciton lifetime by using vibrational strain of a gallium arsenide (GaAs) resonator. The strain-induced modulations have anisotropic dependences on the crystal orientation, which reveals the origin of these modulations to be the piezoelectric effect. Numerical analyses based on the tunneling model clarify that the mechanical strain modulates the internal electric field and spatially separates the electrons and holes, leading to non-radiative exciton decomposition. This carrier separation also generates an optomechanical back-action force from the photon to the resonator. Therefore, these results indicate that the mechanical motion can be self-modulated by exciton decays, which enables one to control the thermal noise of the resonators and provides a photon-exciton-phonon interaction in solid-state systems.

## I. INTRODUCTION

Mechanical strain can be used to control the electronic states of various solid-state systems, such as quantum dots in semiconductors [1-4] and nitrogen-vacancy (NV) centers in a diamond [5-8]. Owing to the deformation potential of the crystals, the energy differences between their excited and ground states change with the mechanical vibration of the resonator, which results in modulation of the photon emission and absorption energies [1-3]. This energy modulation alters the interaction between two-level systems and the mechanical resonator [9], which leads to mechanical mode cooling and lasing [10-12]. Mechanical strain has also been used to control the electron spin states in NV centers [5-7]. This spin-strain interaction enables ones to coherently manipulate a single spin in the strong driving regime and sustain spin coherence against environmental magnetic field fluctuations [7]. The strain field, moreover, breaks the crystal symmetries, which invokes coherent coupling between excitons with different angular momenta, called bright and dark states [13,14]. This strain-induced coupling externally modifies the optical selection rule of the two-level systems and allows for practical usage of the long-lived dark states.

The relaxation of excitons, namely the recombination or decomposition processes of electrons and holes, plays an essential role in opt-electrical and optomechanical interactions. A. Barg et al. have recently revealed how the exciton relaxation affects the optical drive efficiency of mechanical motion and experimentally showed how to optimize it with an external electric field [15]. However, its counter process, how mechanical motion affects the exciton relaxation process, has remained obscure. In this paper, we investigated the lifetime of GaAs bound excitons affected by vibrational strain of a mechanical resonator by making time-resolved stroboscopic photoluminescence (PL) measurements. We experimentally clarified that the mechanical strain modulates the lifetime of the excitons via the piezoelectric field. This strain-induced electric field allowed the electrons to pass through the Coulomb potential generated by the holes and causes decomposition of the excitons depending on the mechanical displacement. Numerical analyses based on this tunneling model reproduced the experimental results and revealed the mechanism of the lifetime and intensity modulations. Moreover, the separated electrons and holes cause a back-action force on the resonator [12,16]; therefore, the lifetime modulation varies the back-action force from the photon to the mechanical motion as a function of the displacement. This self-modulation process provides an optomechanical feedback loop via exciton decays, which enables us to optically control the mechanical motion towards mode cooling and lasing.

## II. SAMPLE AND EXPERIMENT

Figure 1(a) shows a schematic image of the GaAs-based mechanical resonator. The resonator was composed of n-doped Al$_{0.3}$Ga$_{0.7}$As (100-nm-thick), undoped GaAs (400-nm-thick), and an undoped Al$_{0.3}$Ga$_{0.7}$As/GaAs superlattice (10-/10-nm-thick, 5 periods). The length and width of the resonator were 37 and 18 μm. Figure 1(b) shows the band diagram of the resonator along [100]. The modulation-doped structure formed a two-dimensional electron gas (2-DEG) layer between the Al$_{0.3}$Ga$_{0.7}$As and GaAs and caused a potential gradient along [100]. The bound excitons, electron-hole pairs trapped by impurities [17], were localized in the undoped GaAs.

The PL intensities from the bound excitons and vibrational amplitude of the resonator were measured with a stroboscopic PL measurement system, which was composed of a pulse laser, spectrometer, piezo actuator, and interferometer as shown in Fig. 1(c). To investigate the effect of the strain on the exciton lifetime, the pump pulses were synchronized to the vibration of the resonator. By changing the relative phase of the pump pulses, PL spectra and their time evolutions could be measured at the arbitrary timing of the mechanical motion with a charge coupled device (CCD) and an avalanche photodiode (APD). The pump wavelength and the spectral and time resolutions of the measurements were 780 nm, 50 μeV and 34 ps, where the exciton lifetime was much shorter than the mechanical period of 1 μs.

Figure 1(d) shows the PL spectrum of the fabricated resonator at equilibrium. The three peaks originated from acceptor bound excitons whose typical linewidths were 200 μeV. The spectra from the two higher peaks (blue) overlapped each other, so we focused on the lowest peak (red) in the following stroboscopic PL measurements. The mechanical displacement of the resonator was measured with a HeNe laser and Doppler interferometer. Figure 1(e) shows the frequency response, where the resonance frequency and quality factor were about 1.05 MHz and 30,000. The experiments were performed at 8 K at 10$^{-5}$ Pa.

## III. RESULTS

Figure 2(a) shows the stroboscopic PL spectra of the bound excitons in the mechanical resonator whose longitudinal direction was along [0-11]. The mechanical motion periodically modulated the exciton energy and PL intensity. We measured the exciton lifetime at relative phases of 0 and π. The lifetime was modulated by the mechanical motion as shown in Fig. 2(b). The origin of these modulations can be attributed to two strain effects. One is the piezoelectric effect, which generates an internal electric field, called the piezoelectric field. For GaAs, the piezoelectric field is generated along [100] with strain along [011] and [0-11], and the signs of the fields change between the two strain directions [18]. The other is the deformation potential, which modifies the exciton energies according to the strain tensors [14, 19]. The energy shifts caused by the deformation potentials are independent of the strain directions between [011] and [0-11]. Therefore, a comparison of the two mechanical resonators along different crystal orientations, which dominantly generated the strain along [011] and [0-11], should reveal the contributions of the deformation potential and piezoelectric effect to these strain-induced modulations. Figures 2(c) and 2(d) show the stroboscopic PL spectra and the time-resolved PL intensity in the resonator oriented to [011]. The energy modulation in this resonator had the same dependence on the mechanical displacement as in the resonator oriented to [0-11], whereas the intensity and lifetime modulations showed an opposite dependence; strong PL was observed at 0 (π) phase for the resonator oriented to [0-11] ([011]). These results reveal that the energy modulation was caused by the deformation potentials, while the lifetime and intensity modulation were due to the piezoelectric effect. We further found a relation between the PL intensity and lifetime. In both resonators, the PL intensities decreased as the lifetime decreased, which indicates that the piezoelectric field changed the non-radiative decay processes of the excitons.

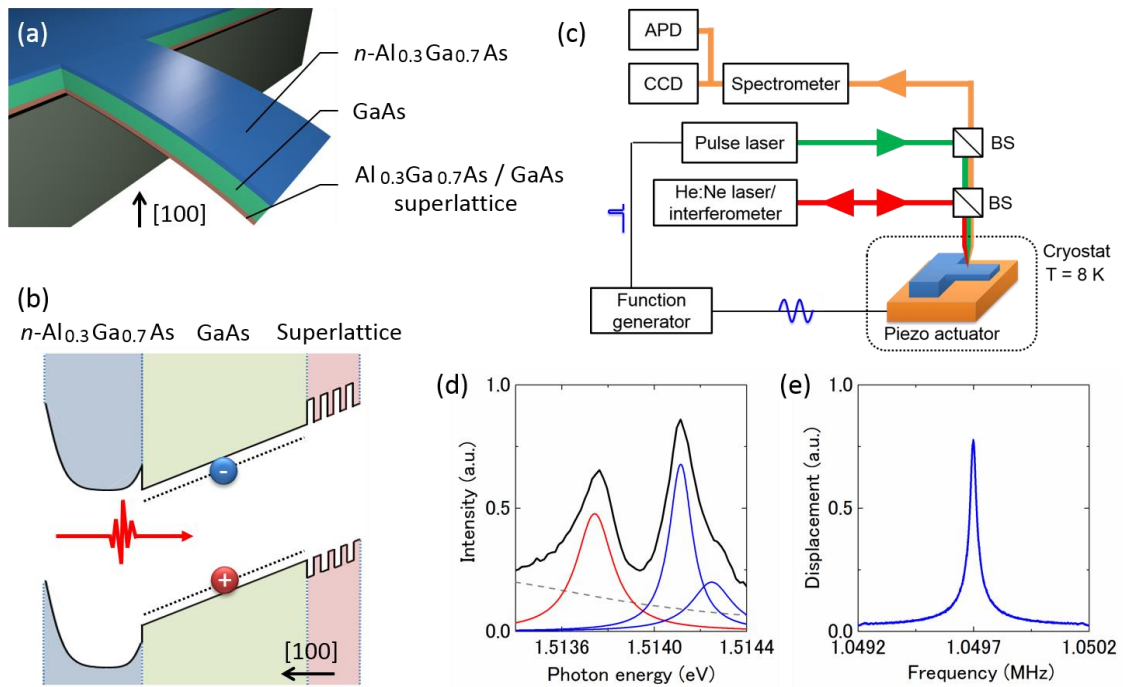

FIG. 1. (a) Schematic image of the mechanical resonator. (b) Band diagram of the resonator along [100]. The modulation-doped structure forms a potential gradient in GaAs. (c) Configuration of the time-resolved stroboscopic PL measurement. The PL spectra and their time-evolution were measured with a CCD and APD, where the pump pulses were synchronized to the vibration. (d) PL spectrum of the GaAs-based resonator at equilibrium. The three peaks (red and blue) come from acceptor bound excitons in GaAs. The dashed line indicates the background signal. (e) Frequency response of the resonator as measured with a He-Ne laser and Doppler interferometer at 8 K at 10$^{-5}$ Pa.

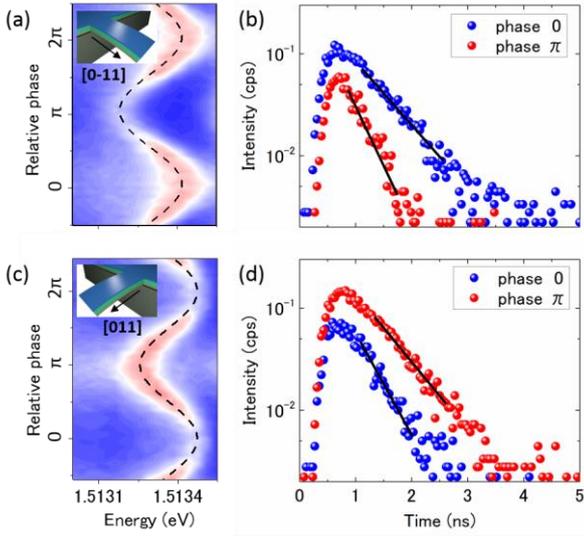

FIG. 2. (a) Stroboscopic PL spectrum of the acceptor bound excitons in the resonator oriented to [0-11]. The vertical axis indicates the relative phase of the pump pulse to the mechanical motion. The inset shows the crystal orientation of the resonator. The dashed line is the fitted curve for the exciton energy. (b) Time evolution of the PL intensity at relative phases of 0 and π. (c), (d) Same as (a) and (b) but for the resonator oriented to [011]. The energy modulations occurred in the same phase, whereas the intensity and lifetime modulations occurred in the opposite phases between the two resonators.

To clarify the mechanism of the intensity and lifetime modulations, we investigated the non-radiative decay processes of the excitons in the presence of built-in and piezoelectric fields. Figure 3(a) shows the band diagram of electrons in GaAs along [100] with a positively charged hole at the center. The potential gradient due to the modulation doping allows the electron to pass through the Coulomb potential generated by the hole. This electron tunneling results in non-radiative decomposition of the excitons, which competes with radiative recombination of the electrons and holes. The overall exciton lifetime ($\tau_{tot}$) and PL intensity ($I_{rad}$) are functions of the tunneling time ($\tau_{tun}$) and radiative decay time ($\tau_{rad}$).

$$\tau_{tot} = \frac{1}{\gamma_{tot}} = \frac{\tau_{rad}\tau_{tun}}{\tau_{rad} + \tau_{tun}} \quad (1a)$$

$$I_{rad} = \frac{\tau_{tun}}{\tau_{rad} + \tau_{tun}} I_0 \quad (1b)$$

Here, $\gamma_{tot}$ and $I_0$ are the decay rate and intrinsic PL intensity of the bound excitons. $\tau_{tun}$ is a function of the internal electric field, which is the sum of the built-in electric field caused by the modulation doping and the piezoelectric field due to the mechanical strain. We numerically calculated $\tau_{tun}$ with the mechanical displacement by using the WKB approximation,

$$\frac{1}{\tau_{tun}} = \gamma_{tun}$$
$$= \frac{\hbar\pi}{8m^* a_B^2} \exp\left(-\frac{2}{\hbar}\int_a^b dz \sqrt{2m^*(V_{(z)} - E_e)}\right) \quad (2a)$$

$$V_{(z)} = -\frac{e^2}{4\pi\varepsilon z} + \left(\epsilon_0 + \epsilon_p \cos(\omega_m t)\right)z \quad (2b)$$

with fitted parameters of $\epsilon_0$ and $\epsilon_p$. $\tau_{rad}$ was assumed to be 650 ps, a typical value for bound excitons in GaAs [20, 21]. $m^*$, $E_e$, and $a_B$ are the effective mass, energy, and Bohr radius of the electron in GaAs, and $\hbar$, $\varepsilon$, $\epsilon_0$, $\epsilon_p$, and $\omega_m$ are the reduced Planck constant, permittivity, built-in electric field, piezoelectric field, and drive frequency of the resonator. Figure 3(b) shows $\tau_{tun}$, $\tau_{tot}$, and $I_{rad}$ calculated for various internal electric fields. $\tau_{tot}$ and $I_{rad}$ exponentially decrease when the internal electric field exceeds 400 V/cm. Figures 3(c) and 3(d) show the experimentally determined and numerically calculated exciton lifetimes and PL intensities in one period of mechanical motion of the resonators oriented to [0-11] (c) and [011] (d). The built-in electric and piezoelectric fields in Fig. 3(c) (3(d)) were 400 (360) and 50 (60) V/cm, respectively. The good agreement between the experimental results and numerical analyses confirmed that the PL intensities and lifetimes of the excitons were modulated by the electron tunneling due to the piezoelectric field.

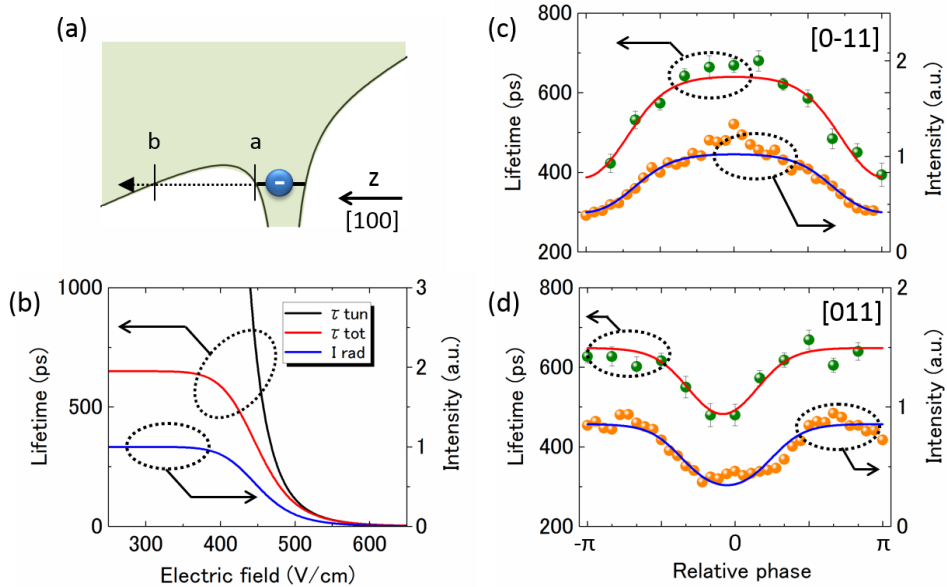

FIG. 3. (a) Energy diagram of an electron in a Coulomb potential of a hole and internal electric field. The electric field causes electron tunneling. (b) Exciton lifetimes and PL intensity calculated with the tunneling model. (c), (d) Experimental (circles) and calculated (solid lines) exciton lifetimes and PL intensity as a function of oscillation phase. The resonators were oriented to [0-11] (c) and [011] (d).

## IV. DISCUSSION

Here, let us discuss the back-action effect on the mechanical motion from the exciton decompositions. According to the results, the internal electric field spatially separates the electrons and holes; the electrons (holes) concentrate at the top (bottom) boundaries of the GaAs. These separated carriers generate an additional electric field along [100], which causes a bending force to be applied to the mechanical resonator via the piezoelectric effect [12, 16]. Therefore, the photon absorption of the bound excitons causes an optomechanical back-action force on the resonator. The intensity of the back-action force ($F$) is the product of the input laser power ($P_{in}$), photon absorption efficiency ($\eta$), carrier separation ratio ($\gamma_{tun}/\gamma_{tot}$), and time-averaged piezoelectric force from one electron-hole pair ($\alpha$).

$$F = \alpha \frac{\gamma_{tun}}{\gamma_{tot}} \eta P_{in} \quad (3)$$

The periodic modulation of $F$ at the resonance frequency excites the mechanical motion. In fact, an amplitude-modulated laser has been used to drive the mechanical vibration [12]. Therefore, the mechanical modulation of $F$, which periodically varies as a function of the displacement, provides a self-modulation of the mechanical motion. This self-modulation enables one to amplify or suppress the mechanical motion via the feedback effect. So far, self-modulation and feedback control of the motion have been demonstrated by using deformation potentials to shift the energies of the excitons, because they modulate $\eta$ in accordance with the mechanical displacement [12]. Besides force modulation by the energy shifts, mechanical modulation of the exciton lifetime by the piezoelectric field will generate an alternative feedback effect because it changes $\gamma_{tun}/\gamma_{tot}$ as a function of the displacement. Figure 4(a) shows the change in the back-action force due to the lifetime modulation in the resonator oriented to [0-11].

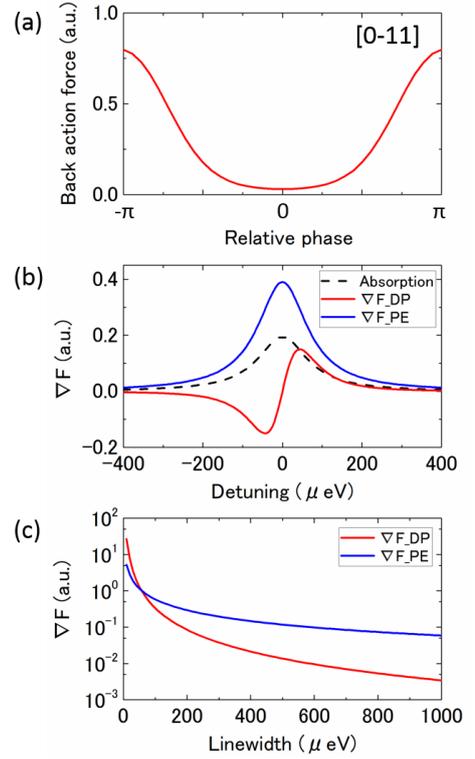

FIG. 4. (a) Mechanical modulation of the back-action force via lifetime modulation in the resonator oriented to [0-11]. (b) Detuning energy dependences of force modulations by the deformation potential ($\nabla F_{DP}$) and piezoelectric effect ($\nabla F_{PE}$). The dashed line indicates the absorption spectrum of the exciton. (c) Linewidth dependences of the force modulations. $\nabla F_{PE}$ ($\nabla F_{DP}$) becomes dominant when the linewidth is wider (narrower) than 60 µeV.

The contributions of the two feedback effects were investigated by calculating $\nabla F = \partial F/\partial z'$, where $z'$ is mechanical displacement. In this comparison, we assumed that $\eta$ has a Lorentzian photon-energy dependence with an exciton linewidth ($\gamma_{ex}$) of 18 GHz. $\nabla F$ is described in terms of the deformation potentials ($\nabla F_{DP}$) and piezoelectric effect ($\nabla F_{PE}$):

$$\nabla F_{DP} = \alpha \frac{\gamma_{tun}}{\gamma_{tot}} \frac{\partial \eta}{\partial z'} P_{in}$$
$$= \alpha \frac{\gamma_{tun}}{\gamma_{tot}} C_{def} \frac{2\delta\omega\gamma_{ex}}{(\delta\omega^2 + \gamma_{ex}^2)^2} P_{in} \quad (4a)$$

$$\nabla F_{PE} = \alpha \frac{\partial}{\partial z'} \frac{\gamma_{tun}}{\gamma_{tot}} \eta P_{in}$$

$$= \alpha \left(\frac{\gamma_{rad}}{\gamma_{tot}^2}\right) \frac{\partial \gamma_{tun}}{\partial \epsilon} C_{piez} \frac{\gamma_{ex}}{\delta\omega^2 + \gamma_{ex}^2} P_{in}$$

(4b)

Here, $C_{def} (= \partial E/\partial z')$, $\delta\omega$, $\epsilon$ and $C_{piez} (= \partial \epsilon/\partial z')$ are the deformation potential of GaAs, detuning energy of the excitation laser and exciton absorption, internal electric field, and piezoelectric constant. We numerically calculated the detuning dependence of $\nabla F_{DP}$ and $\nabla F_{PE}$ in the resonator oriented to [0-11] with the experimentally obtained $C_{def}$ and $C_{piez}$, as shown in Fig. 4(b). $\nabla F_{DP}$ ($\nabla F_{PE}$) is an odd (even) function of the detuning energy of the excitation laser. The maximum $\nabla F_{PE}$ is two times larger than that of $\nabla F_{DP}$ in this resonator. Figure 4(c) shows the linewidth dependences, which indicates that $\nabla F_{PE}$ is less affected by linewidth broadening. In this resonator, $\nabla F_{PE}$ becomes the dominant mechanism of the self-modulation when the linewidth is wider than 60 µeV. This indicates that the lifetime modulation scheme has the potential to become useful in systems, whose exciton linewidths are too wide for conventional absorption modulation schemes, such as quantum wells [22, 23] and two-dimensional materials [24, 25].

## V. CONCLUSION

We mechanically modulated the exciton lifetimes and luminescence intensities with the vibrational strain and piezoelectric effect. The numerical calculations based on the tunneling model clarified that the piezoelectric field modulates the decomposition rate of electrons and holes. The results furthermore indicate an alternative optomechanical force modulation mechanism due to the exciton relaxations that enables one to manipulate the mechanical motion by photon absorption. This carrier-mediated optomechanical interaction will be an intermedium of photons, photons, and electrons and will pave the way to hybridizing nanoscale and macroscopic physics.

## ACKNOWLEDGMENTS

We thank M. Asano, R. Okuyama, and S. Adachi for their discussions on the theoretical models and experimental results. This work was supported by MEXT KAKENHI Grants (Nos. JP15H05869 and JP16H01057).